\documentclass[amsmath,nobibnotes,aps,superscriptadress,amssymb,pra,aps,showpacs,superscriptaddress,twocolumn, longbibliography]{revtex4-1}
\usepackage{amsmath,amsfonts,amssymb,amsthm,graphics,graphicx,epsfig,bbm}
\usepackage[colorlinks=true,citecolor=blue,linkcolor=blue,urlcolor=blue]{hyperref}
\usepackage[usenames]{color}
\usepackage{graphicx}
\usepackage{subfigure}
\usepackage{amsmath}
\usepackage{tikz}
\usetikzlibrary{quantikz}
\usepackage{epsfig}
\usepackage{dcolumn}
\usepackage{bm}
\usepackage{color}
\usepackage{times}
\usepackage{epstopdf}
\usepackage{amssymb}
\usepackage{amstext}
\usepackage{latexsym}
\usepackage{float}
\usepackage{hyperref}
\usepackage{amsfonts}
\usepackage{psfrag}
\usepackage{soul,xcolor}
\usepackage[normalem]{ulem}
\usepackage{dsfont}
\usepackage{txfonts}
\usepackage{physics}
\usepackage{footnote}
\usepackage{multirow}
\usepackage{appendix}
\usepackage{mathtools}
\usepackage{xspace}

\newtheorem*{theorem*}{Theorem}

\usepackage{mathtools}

\begin{document}

 \title{Loschmidt echo and scrambling of systematic errors  in tomography - a quantum signature of chaos}

	\author{Abinash Sahu}
\email{abinashsahu96@gmail.com}
\affiliation{Department of Physics, Indian Institute of Technology Madras, Chennai, India, 600036}
\affiliation{Center for Quantum Information, Communication and Computing,
	Indian Institute of Technology Madras, Chennai, India 600036}
\author{Naga Dileep Varikuti}
\email{dileep.varikuti@unitn.it}
\affiliation{Department of Physics, Indian Institute of Technology Madras, Chennai, India, 600036}
\affiliation{Center for Quantum Information, Communication and Computing,
	Indian Institute of Technology Madras, Chennai, India 600036}
\affiliation{Pitaevskii BEC Center, CNR-INO and Department of Physics, University of Trento, Via Sommarive 14, I-38123 Trento, Italy}
\affiliation{INFN-TIFPA, Trento Institute for Fundamental Physics and Applications, Via Sommarive 14, I-38123 Trento, Italy}
\author{Vaibhav Madhok}
\thanks{corresponding author}
\email{madhok@physics.iitm.ac.in}
\affiliation{Department of Physics, Indian Institute of Technology Madras, Chennai, India, 600036}
\affiliation{Center for Quantum Information, Communication and Computing,
	Indian Institute of Technology Madras, Chennai, India 600036}



\begin{abstract}
{How does quantum chaos lead to rapid scrambling of information as well as {systematic} errors across a system when one introduces perturbations in the dynamics? What are its consequences for the reliability of quantum simulations and quantum information processing?
We employ continuous measurement quantum tomography as a paradigm to study these questions. The measurement record is generated as a sequence of expectation values of a Hermitian observable evolving under repeated application of the Floquet map of the quantum kicked top. We construct a quantity to capture the scrambling of {systematic} errors, an out-of-time-ordered correlator (OTOC),  that serves as a signature of chaos and quantifies the spread of errors. We show that the spread of errors, as quantified by the OTOC, are related to the operator Loschmidt echo (OLE) which is defined as the Hilbert-Schmidt inner product of the operators $\mathcal{O}_n$, and $\mathcal{O'}_n$ generated from repeated application of the Floquet map for ideal (unperturbed) dynamics and the
\emph{true} (perturbed) dynamics,  respectively. 
This also gives us an operational interpretation of Loschmidt echo for operators by connecting it to the performance of quantum tomography. {We show how our results demonstrate not only a link between LE and scrambling of errors different than previous studies, but that such a link can have operational consequences in quantum information processing.}}
\end{abstract}
\maketitle

\section{Lead paragraph}
The connections between information, complexity and chaos are at the foundations of statistical mechanics both in classical as well as quantum mechanics. These are now being actively pursued as they have relevance to quantum information processing. 
We study footprints of chaos in the quantum world 
as well as its consequences to quantum simulation of many-body Hamiltonians under the influence of systematic errors. We use quantum tomography as a paradigm to study how chaotic dynamics in the presence of {systematic errors} leads to information gain of an unknown quantum state. {Interestingly, this has connections to metrics of complexity like the out of time ordered correlator (OTOC). The OTOC captures the incompatibility of two operators when one of them is subject to time evolution in the Heisenberg picture. The OTOCs capture the information ``scrambling" or the growth of complexity of initially localized operators due to system dynamics. 
The OTOCs we construct capture the incompatibility of operators arising due to systematic errors  in the Hamiltonian and we connect this to another widely studied concept to characterize quantum chaos, the Loschmidt echo, in a novel way different from previous studies.} 

\section{Introduction}

The central goal of quantum chaos is to inform us about the properties of a quantum system whose classical counterpart is chaotic. How does chaos manifest in the quantum world, and what notions of complexity might be suitable to quantify it? The state vector of a system in quantum theory resides in Hilbert space, which is a big space~\cite{caves1996quantum}. Quantum theory permits the state of the system to be any vector in this space, even permitting a coherent superposition of possibilities considered mutually exclusive in the classical world. Therefore, while classically chaotic dynamics generates classical information in the form of complex classical trajectories, quantum chaotic dynamics generate quantum information in the form of pseudo-random vectors in the Hilbert space. These states typically have a high entropy which is calculated according to a fixed fiducial basis \cite{wootters1990random, bengtsson2017geometry}. {Vigorous thrust in the understanding of quantum many-body dynamical systems through dynamically generated entanglement ~\cite{miller1999signatures, bandyopadhyay2002testing, wang2004entanglement, trail2008entanglement, furuya1998quantum, lakshminarayan2001entangling, seshadri2018tripartite} and quantum correlations~\cite{madhok2015signatures,madhok2018quantum}, deeper studies in the ergodic hierarchy of quantum dynamical systems~\cite{gomez2014towards, bertini2019exact, aravinda2021dual, vikram2022dynamical} have marked important advances in the field. These, coupled with the traditional approach to studies of level statistics~\cite{haake1991quantum} and  Loschmidt echo (LE)~\cite{peres1984stability, peres1997quantum, Goussev2012loschmidt, gorin2006dynamics} and complemented by the ability to coherently control and manipulate many-body quantum systems in the laboratory~\cite{gong2001coherent, gong2005quantum, brif2010control, smith2013quantum, mirkin2021quantum}, have brought us to a fork in our path. On the one hand, this is a harbinger of the possibility of building quantum simulators, an important milestone in our quest for the holy grail- a many-body quantum computer. On the other hand, the same properties that make quantum systems generate complexity will make them sensitive to errors that naturally occur in implementing many-body Hamiltonians.

{Out-of-Time-Ordered correlators (OTOCs) capture the operator growth and scrambling of quantum information and have been very useful as a probe for chaos in quantum systems~\cite{maldacena2016bound, swingle2016measuring, hashimoto2017out, kukuljan2017weak, swingle2018unscrambling, wang2021quantum, sreeram2021out, varikuti2022out}.}}  {The OTOCs, originally introduced in the context of superconductivity~\cite{larkin}, have been widely used as diagnostics of information scrambling~\cite{ope2, ope1, ope4, ope5, lin2018out, shukla2022out} and quantum chaos~\cite{pawan, seshadri2018tripartite, lakshminarayan2019out, shenker2, moudgalya2019operator, omanakuttan2019out, manybody2, prakash2020scrambling, prakash2019out, varikuti2022out, markovic2022detecting, dileep2024, martinez2023stochastic}, and have also been explored in studies of many-body localization~\cite{manybody3, manybody4, manybody1, huang2017out} and holographic systems~\cite{shock1, shenker3}. Given two fiducial Hermitian and/or unitary operators $W$ and $V$, the out-of-time-ordered commutator function in a given quantum state $\rho$ can be computed as follows:}

{\begin{eqnarray}\label{commutator}
C_{\rho, \thinspace WV}(t)=\frac{1}{2}\text{Tr}\left(\rho \left[W(t), V\right]^{\dagger}\left[W(t), V\right]\right),
\end{eqnarray}}
{where $W(t)=\hat{U}^{\dagger}(t)W\hat{U}(t)$ denotes the Heisenberg evolution of the operator $W$ under the evolution generated by the system Hamiltonian. For the numerical and the experimental perspectives it is often convenient to consider $\rho$ to be maximally mixed, i.e., $\rho=\mathbb{I}/d$, where $d$ is the Hilbert space over which $\rho$ is supported. When $W$ and $V$ are taken to be Hermitian, the commutator function in Eq. (\ref{commutator}) contains a two-point and a four-point correlator:}
{\begin{eqnarray}
 C_{WV}= \dfrac{1}{d}\left[ \text{Tr}\left(W^2(t)V^2\right)-\text{Tr}\left( W(t)VW(t)V \right) \right].  
\label{OTOC}
\end{eqnarray}}
{Note that we have omitted $\rho$ from the subscript on the left-hand side of the above expression. Due to the unusual time ordering, the four-point correlator is usually referred to as the OTOC. This is because the behavior of $C_{WV}(t)$ depends predominantly on the four-point correlator. Therefore, the terms commutator function and OTOC are frequently used interchangeably to refer to the same quantity, $C_{WV}(t)$.  Scrambling is captured in the commutator $ C(t)=-\langle[W(t),V(0)]^2\rangle$ between two Hermitian operators $V$ and $W$. If the two operators are local and separated (local operators at different sites of a $1D$ spin chain for example), initially the commutator is zero, since the operators are acting on independent subspaces. The time at which the commutator becomes appreciably divergent from zero is termed as scrambling time. In this work, we construct an analogus quantity, which we call the \textit{error OTOC}, that captures the scrambling of systematic errors across the system.}

{The sensitivity to perturbations in a quantum system is usually quantified by LE which is defined as the fidelity, $F(\tau)=\lvert\bra{\psi_0} e^{iH'\tau/\hbar}e^{-iH\tau/\hbar}\ket{\psi_0}\rvert^2,$ where $\ket{\psi_0}$ is the state undergoing evolution for a time $\tau$ and $H$ and $H'$ are the unperturbed and perturbed Hamiltonians respectively.  
 In the absence of perturbations, LE attains the maximum  value of one. The presence of imperfections leads to a decay. The connection between echo dynamics and decoherence effects has been studied and the former has been used to quantify the latter \citep{zurek2001sub,cucchietti2003decoherence,cucchietti2004universality}. 
LE is studied in various  fields of physics, including quantum chaos \cite{ Goussev:2012, gorin2006dynamics}, quantum computation \cite{georgeot2001stable,georgeot2002quantum,frahm2004quantum,garcia2008shor} and quantum phase transition \cite{andraschko2014dynamical,zheng2008loschmidt,vanhala2023theory}. In our work, we show numerically how the sensitivity to systematic errors in the Hamiltonian, as captured by LE, behaves analogous to the error scrambling as captured by the error OTOC  as one varies the degree of chaos in the dynamics. This establishes a connection between a measure of operator incompatibility and scrambling to LE.  Our work links two quantifiers of quantum chaos in an operational way that can be explored using current experimental techniques in the setting of cold atoms interacting
with lasers and magnetic fields.}

\begin{figure*}[htbp]
  	\centering
    \includegraphics[scale=0.250]{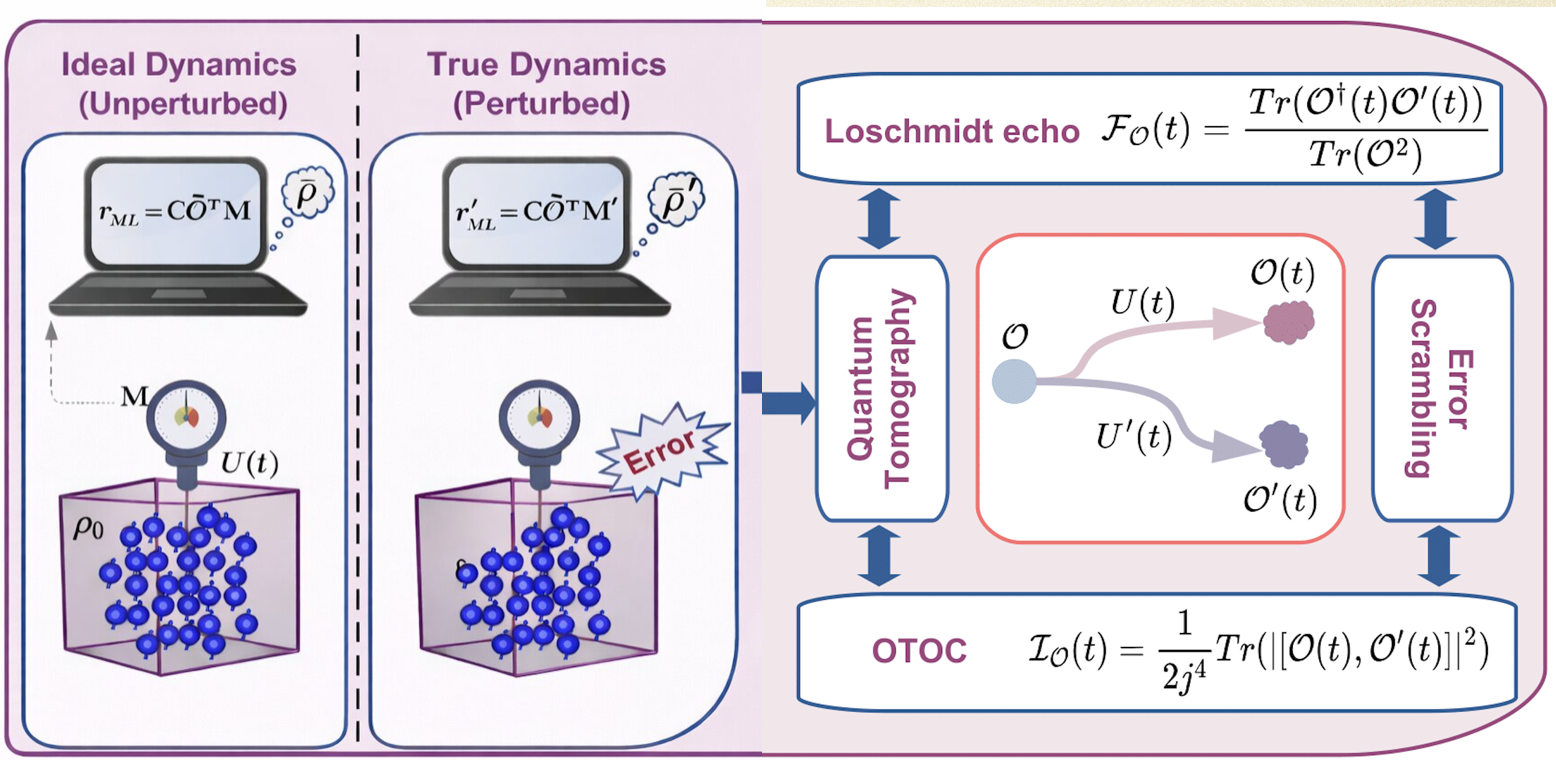}
  	
  	\caption {An illustration of continuous measurement tomography and its connection to various quantifiers of quantum chaos. In the ideal scenario where the experimentalist has complete knowledge of the dynamics, the reconstructed state reaches the actual state with time. However, in reality, we have errors and the experimentalist is ignorant about the true/actual (perturbed) dynamics that leads to an improper reconstruction of the quantum state. Thus, the reconstruction fidelity decays after some time and positively correlates with the operator Loschmidt echo. We quantify the scrambling of error as $\mathcal{I}_{\mathcal{O}}(t)$, the error OTOC, and connect it with LE given by $\mathcal{F}_{\mathcal{O}}(t)$.}
  	\label{fig:graphical}
  \end{figure*}

{Quantum tomography gives us a window to study sensitivity to errors in quantum  chaotic Hamiltonians~\cite{lloyd1996universal,johnson2014what} and establish the connections discussed above. Quantum tomography uses the statistics of measurement records on an ensemble of identical systems in order to make the best estimate of the actual state $\rho_0$. {We consider continuous weak measurement tomography protocol~\cite{silberfarb2005quantum,smith2006efficient,riofrio2011quantum,smith2013quantum, PhysRevLett.95.030402, Riofrio2011, PhysRevA.90.032113, PhysRevLett.121.130404, PhysRevLett.97.180403, PhysRevA.87.030102, PhysRevLett.111.160401, PhysRevLett.114.090403, hacohen2016quantum, PhysRevLett.115.180407}, and the time series of operators can be generated by the Floquet map of a quantum dynamical system to investigate the role of chaos on the information gain in tomography~\cite{madhok2014information,sreeram2021quantum,sahu2022effect, PhysRevB.108.224306}, quantum simulations \cite{PhysRevLett.124.230501} and quantum phase transitions \cite{PhysRevA.106.032215}. {Continuous weak measurements have been employed 
to address foundational questions like quantum-to-classical transition and emergence of classical chaos from underlying quantum mechanics \cite{habib2006emergence, bhattacharya2000continuous, arora2025quantum}}.
Here, we connect two quantifiers of quantum chaos, namely LE and error OTOC, through continuous weak measurement tomography as illustrated in Fig. \ref{fig:graphical}.} Our approach, besides establishing a link between LE and scrambling of errors as captured by error OTOCs, can be tested in the laboratory. Experimentally, it's a huge challenge to measure information scrambling. For example, the standard measurement of OTOCs will involve a backward evolution in time as evident from Eq. \ref{OTOC}. Similarly, LE, by definition, requires a backward ``echo" or evolution.  Our proposal to study chaos and scrambling using continuous measurement tomography needs only single shot forward evolutions of the system with simple modification to the existing experiments.}


{There is an interesting analogy in our approach and the notion of chaos as captured by the positive Kolmogorov-Sinai (KS) entropy \cite{pesin1977characteristic}.  Chaos, classically as well as in quantum mechanics, has an intimate connection with information gain. Chaos implies unpredictability. Time-evolved trajectories twist and wind away from each other at an exponential rate and then fold back to remain confined in a bounded phase space respecting ergodicity.
The flip side of this unpredictability  is the potential information that can be obtained if one tracks these trajectories. The perspective is quantified by the KS entropy \cite{pesin1977characteristic}, which is defined as the rate of information gain, at increasingly fine scales, about the missing information in classical chaos - \emph{the initial conditions}~\cite{caves1997unpredictability}. KS entropy is equal to the sum of positive Lyapunov exponents in the system and hence, a positive KS entropy implies chaos. 
Extending this to the quantum domain, we connect the quantum chaotic dynamics to the information gain about the \emph{the initial conditions} - which is exactly the task of quantum tomography. We must however mention that this analogy between classical and quantum chaos has a natural limitation. While classically it is possible to generate an infinite amount of information and determine the missing information about the \textit{initial conditions} at increasingly finer scales, one can only extract a limited amount of information through quantum measurements for a finite dimensional quantum system.  A given $N$ dimensional quantum system can only yield $O(N)$ bits of information. A single qubit, for example, can only yield one bit of classical information.}

While chaotic dynamics is a source of information quantified by a positive KS entropy, it is sensitive to errors, as captured by LE and error scrambling. In many body systems, quantum or classical, we must expect the presence of both chaos and errors. In this work, we address this scenario; we go on to discover quantum signatures of chaos while shedding light on the larger question of many-body quantum simulations under \textcolor{blue}{systematic} perturbations. While the KS entropy enables a rapid information gain, LE will cause a rapid accumulation of errors as we quantify using error OTOCs. This interplay between KS entropy and LE is a generic feature of any many-body system, and we identify and quantify the crossover between these two competing effects.

\section{Continuous measurement tomography} 

{In this section, we briefly outline the details concerning continuous measurement tomography. }An ensemble of $N$ identical systems, initially prepared in the product state $\rho^{\otimes N}_0$ undergoes a separable time evolution by a unitary $U(t)$. A weakly coupled probe will generate the measurement record by performing weak continuous measurement of an observable $\mathcal{O}$. For sufficiently weak coupling, the randomness of the measurement outcomes is dominated by the quantum noise in the probe rather than the measurement uncertainty, i.e., the projection noise. In this case, the quantum backaction is negligible, and the state remains approximately separable. Thus, we get the stochastic measurement record 
\begin{equation}
	M(t)=\mathrm{Tr}(\mathcal{O}(t)\rho_0)+W(t),
	\label{tom_noise}
\end{equation}
where $\mathcal{O}(t)=U^{\dag}(t)\mathcal{O}U(t)$ is the time evolved operator in Heisenberg picture, and $W(t)$ is a Gaussian white noise with spread $\sigma/N$. 

Any density matrix of Hilbert-space dimension $d$ can be realized as a generalized Bloch vector $\bf r$ by expanding $\rho_0=I/d+\Sigma^{d^2-1}_{\alpha=1}\ r_\alpha E_\alpha$ in an orthonormal basis of traceless Hermitian operators $\{E_\alpha\}$. We consider the measurement record at discrete times as $M_n=M(t_n)=Tr(\mathcal{O}_n\rho_0)+W_n$, that allows one to express the measurement history 
\begin{equation}
	{\bf M}=\tilde{\mathcal{O}}{\bf r}+{\bf W},
	\label{ms}
\end{equation}
where $\tilde{\mathcal{O}}_{n\alpha}=\tr(\mathcal{O}_{n}E_\alpha)$. Thus, the problem of quantum tomography is reduced to linear stochastic state estimation of $\rho_0$ given $\{M_n\}$.
In the limit of negligible backaction, the probability distribution associated with measurement history $\bf M$ for a given state vector $\bf r$ is \cite{silberfarb2005quantum,smith2006efficient}
\begin{equation} 
	\begin{split}
		p({\bf M|r}) & \varpropto \mathrm{exp}\ \Big\{-\frac {N^2}{2\sigma^2}\sum_{i}[M_i-\sum_{\alpha}\tilde{\mathcal{O}}_{i\alpha}r_\alpha]^2\Big\}
		\\
		& \varpropto \mathrm{exp}\ \Big\{-{\frac {N^2}{2\sigma^2}\sum_{\alpha,\beta}({\bf r-r_{ML}})_\alpha\ C^{-1}_{\alpha\beta}\ ({\bf r-r_{ML}})_\beta\Big\}}.
		\label{pdf_uniform_prior}
	\end{split}
\end{equation}
In the weak backaction limit, the fluctuations around the mean are Gaussian distributed, and hence the maximum likelihood estimate of the Bloch vector components is the least-squared fit as
\begin{equation}
	{\bf r}_{ML}=\bf C\tilde{\mathcal{O}}^{T}M,
	\label{rml1}
\end{equation}
where ${\bf C=(\tilde{\mathcal{O}}^T\tilde{\mathcal{O}}})^{-1}$ is the covariance matrix and the inverse is Moore-Penrose pseudo inverse~\cite{ben2003generalized} in general. The estimated Bloch vector ${\bf r}_{ML}$ may not always represent a physical density matrix with nonnegative eigenvalues due to the finite signal-to-noise ratio. Therefore, we impose the constraint of positive semidefiniteness on the reconstructed density matrix and obtain the physical state closest to the maximum-likelihood estimate. To do this, we employ a convex optimization procedure~\cite{vandenberghe1996semidefinite} where the final estimate of the Bloch vector $\bf \bar{r}$ is obtained by minimizing the argument 
\begin{equation}
	||{\bf r}_{ML}-{\bf \bar{r}}||^2=({\bf r}_{ML}-{\bf \bar{r}})^T{\bf C}^{-1}({\bf r}_{ML}-\bf \bar{r})
\end{equation}
subject to the constraint $$I/d+\Sigma^{d^2-1}_{\alpha=1}\ \bar{r}_\alpha E_\alpha\geq0.$$\\


	

The above description of the continuous weak measurement tomography provides us with a  paradigm for probing error scrambling as captured by OTOCs, LE and information gain in a single shot experiment as opposed to tomography using projective measurements. 
{More importantly, our reason for exploring quantum chaos within this paradigm is that weak measurement protocols allow information to be extracted while keeping the evolution of the collective ensemble close to unitary (negligible back-action regime), and remain experimentally feasible.}
Other measurement schemes, like projective measurements, will collapse the state and cause back-action and therefore may not be suitable to explore the information generation properties of chaotic unitary operators. 

\section{Tomography with imperfect knowledge}
{The above description of the continuous measurement tomography protocol corresponds to an idealized scenario in which the experimentalist has complete knowledge of the dynamics. The ideal dynamics are represented by unprimed variables, including the time-evolved observables $\mathcal{O}_n$ and the covariance matrix $\mathbf{C}$, from which the state can be accurately reconstructed using the procedure described in the previous section.}
However, in reality, one never knows the actual/true underlying dynamics, and there is always a departure from the ideal case due to inevitable errors and the true dynamics carries perturbations. Thus, the experimentalist, oblivious to these errors, models their estimation using a covariance matrix, ${\bf C=(\tilde{\mathcal{O}}^T\tilde{\mathcal{O}}})^{-1}$, assuming the ideal scenario. However, the measurement record, $M'$ arises from the true/actual dynamics denoted by the primed variables. As a result they end up reconstructing an incorrect state $\bar{\rho}'$ from
\begin{equation}
	{\bf r'}_{ML}=\bf C\tilde{\mathcal{O}}^{T}M'.
	\label{rml2}
\end{equation}
In the above equation, the measurement record is obtained from the measurement device (probe), and the experimentalist is ignorant about the \emph{actual/true} dynamics (which is accompanied by perturbations relative to the idealized dynamics as assumed by the experimentalist), given by the unitary $U'(t)$, that has generated this record. Hence, the covariance matrix is  determined from the experimentalist's version of the dynamics given by the unitary $U(t)$ and the initial observable $\mathcal{O}$. Thus, the ignorance about the error in the dynamics directs the operator trajectory away from the actual one, leading to an improper reconstruction of the state $\rho_0$.


\begin{figure*}[htbp]
	\centering
	\includegraphics[scale=0.56]{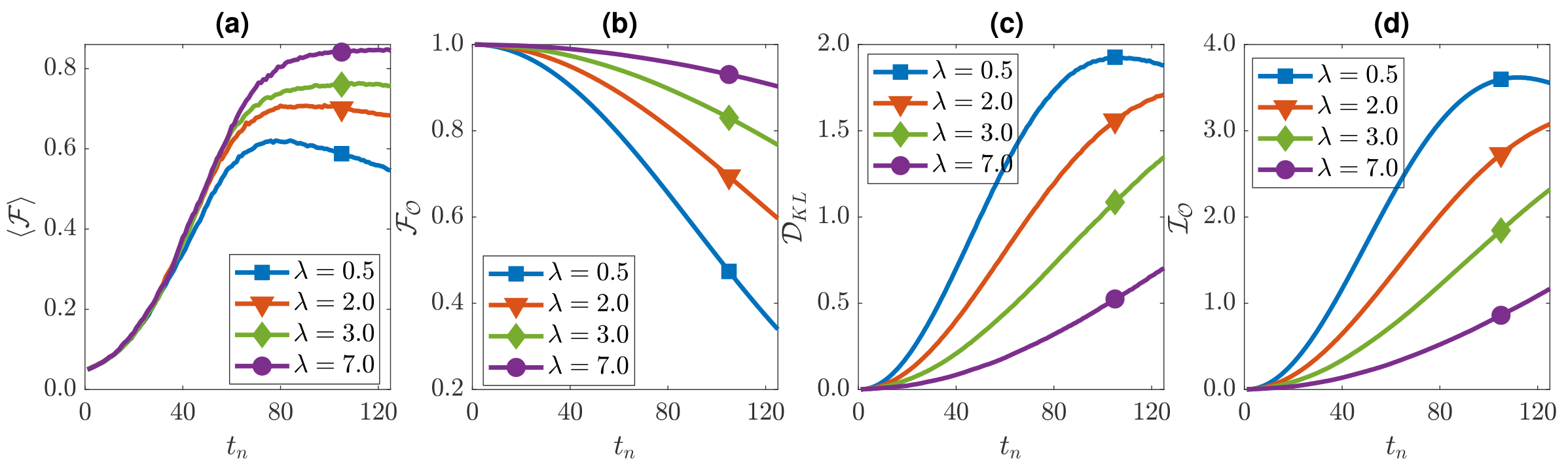}
	
	\caption {{Effect of perturbation on tomography quantified by different metrics as a function of time with an increase in the level of chaos. The kicked top Floquet map of \emph{true} (perturbed) dynamics $U'_\tau=e^{-i(\lambda+\delta\lambda) J^2_z/2J}e^{-i\alpha J_x}$ with $\delta\lambda=0.01$, and ideal (unperturbed) dynamics $U_\tau=e^{-i\lambda J^2_z/2J}e^{-i\alpha J_x}$ generate the time series of operators for a spin $j=10$ and fixed $\alpha=1.4$.} (a) Average reconstruction fidelity $\langle\mathcal{F}\rangle$ of the state $\bar{\rho}'$ derived from Eq. (\ref{rml2}) relative to the actual state $\ket{\psi_0}$, where the average is taken over 100 Haar random states. (b) The operator Loschmidt echo $\mathcal{F}_{\mathcal{O}}$ between two operators. (c) The quantum relative entropy $\mathcal{D}_{KL}$ of the regularized operator evolved under unperturbed dynamics to the operator evolved under perturbed dynamics.   (d) The operator incompatibility $\mathcal{I}_{\mathcal{O}}$, captured by the error OTOC, quantifies the scrambling of errors.}
	\label{fig:fid1}
\end{figure*}


Our goal is to study the effect of the perturbation on the information gain in tomography in the presence of chaos. To accomplish this, we implement the quantum kicked top~\cite{haake1987classical,haake1991quantum,chaudhury2009quantum} described by the Floquet map $F_{QKT}=e^{-i\lambda J^2_z/2J}e^{-i\alpha J_x}$ as the unitary for a period $\tau$ ({that can be set to unity}) , and the unitary at $n^{\mathrm{th}}$ time step is $U(n\tau)=U^n_{\tau}$. Note that the measurement record generated by such periodic application of the Floquet map is not informationally complete, as it leaves out a subspace of dimension $\geq d-2,$  out of $d^2-1$ dimensional operator space \cite{merkel2010random}. 
For our current work, we fix the linear precision angle $\alpha=1.4$ and choose the kicking strength $\lambda$ as the chaoticity parameter. As $\lambda$ is varied from $0$ to $7$, the underlying classical dynamics transitions from a highly regular regime to fully developed chaos. The dynamics that represents the \emph{true} evolution is perturbed relative to the idealized dynamics given by $F_{QKT}$, and we choose a small variation in the kicking strength, $\lambda+\delta\lambda$, and the perturbed unitary becomes $U'_{\tau}=e^{-i(\lambda+\delta\lambda) J^2_z/2J}e^{-i\alpha J_x}$. For our analysis, we consider the dynamics of the quantum kicked top for a spin $j=10$, and perturbation strength $\delta\lambda=0.01$. {Throughout this work, we have considered a systematic error to the dynamics. The randomization of a systematic error is interestingly a way to cancel noise in analog quantum information protocols and paves a way to make these more robust to random noise. These are very interesting directions for the future.}

{Does having more chaos aid or inhibit quantum state reconstruction? 
{As observed previously, the reconstruction fidelity of spin coherent states decreases with increasing chaos~\cite{sahu2022effect}.} This contrasts with results connecting information gain in tomography of random states with the degree of chaos in the dynamics that drives the system \cite{PhysRevLett.112.014102, madhok2016characterizing, PhysRevA.104.032404}. {These findings suggest that} the connection between chaos and information gain depends on the localization properties of the state, i.e., their inverse participation ratio, the degree of chaos, and how well the state is aligned with time-evolved measurement observables \cite{sahu2022effect}.
{Therefore, to study the effect of the degree of chaos on the performance of noisy tomography purely, we consider random initial states measured via random initial observables (generated by rotating $J_x$ through a random unitary) picked from the appropriate Haar measure.}}
We apply our reconstruction protocol on an ensemble of 100 random pure states sampled from the Haar measure on SU$(d)$, where $d=2j+1=21$. We choose one random initial observable and generate the measurement record from the repeated application of the Floquet map of the quantum kicked top. The fidelity of the reconstructed state $\bar{\rho}'$ obtained from Eq. (\ref{rml2}) is determined relative to the actual state $\ket{\psi_0}$, $\mathcal{F}=\bra{\psi_0}\bar{\rho}'\ket{\psi_0}$ as a function of time. {As shown in Fig.~\ref{fig:fid1}a (see also the following section),} the reconstruction fidelity increases in the beginning despite the errors, and after a certain period of time, it starts decaying. The rise in fidelity during the initial time period is because any information, even if partially inaccurate, about a completely unknown random state offsets the presence of errors in its estimation. However, as time progresses, the effect of errors becomes significant. Beyond a certain time, we observe a decline in fidelity as the dynamics continues to accumulate errors that dominate the archive of information present in the measurement record. Most interestingly, the rate of this fidelity decay is inversely correlated with the degree of chaos in the dynamics. 
{This can be viewed as an interplay between rapid information gain due to Lyapunov divergence---a ``quantum" analogue of the classical KS entropy, and the LE, which leads to error accumulation that cause fidelity decay.}

%
%
{In the presence of errors, the dynamics not only generates information at a finite rate but can also exhibit strong sensitivity to those errors, depending on the degree of chaos in the system.} In the supplementary section, we detail the behavior of the role of noise on information gain for a map with fixed chaoticity.

\section{ Quantum signatures of chaos in noisy tomography}

We now quantify the role of chaos in tomography when the error in the dynamics influences our ability to reconstruct the random quantum states. It is evident from Fig. \ref{fig:fid1}a that the rate of drop in fidelity decreases with an increase in the strength of chaos for small perturbations in the dynamics. 
{To elucidate the foregoing discussion, we consider several metrics that exhibit a remarkable correlation with the reconstruction fidelity shown in Fig.~\ref{fig:fid1}a.} First, we define the operator Loschmidt echo (OLE) $\mathcal{F_O}$ as the Hilbert-Schmidt inner product of the operators $\mathcal{O}'_n$, and $\mathcal{O}_n$ generated from repeated application of the Floquet map for \emph{true} (perturbed) dynamics $U'_\tau$ and ideal (unperturbed) dynamics $U_\tau$ of the kicked top on the initial observable $\mathcal{O}$
\begin{equation}
	\mathcal{F_O}(t_n)=\frac{\tr(\mathcal{O}^{\dagger}_n\mathcal{O}'_n)}{\tr(\mathcal{O}^2)}.
\end{equation}
{Analogous to the LE for states, the OLE quantifies the overlap between the operators $\mathcal{O}_n$ and $\mathcal{O}'_n$ and is therefore expected to decay with time. We confirm this behaviour from Fig.~\ref{fig:fid1}b. The figure illustrates} that the OLE decays much slower when the dynamics is chaotic than when it is regular. This behavior correlates positively with the rate of decrease in reconstruction fidelity, as demonstrated in Fig. \ref{fig:fid1}a. The greater the distance between the operators at a given time, the greater the difference between the expectation values with respect to the state and the archive of the measurement record obtained through the time series. {Thus}, our results give an \textit{operational interpretation} of the OLE by connecting it to a concrete physical task of continuous measurement quantum tomography. This also points to a beneficial way to probe these quantities in experiments using current techniques.

Another useful metric is the quantum relative entropy - a measure of distance between two quantum states. Here, we use this metric to measure the distance between two operators $\mathcal{O}_n$ and $\mathcal{O}'_n$. To treat both observables as density operators, we regularize them as follows. We construct a positive operator from an observable by retaining its eigenvectors and taking the absolute value of its eigenvalues. To normalize this operator, we divide it by its trace.
While calculating the relative entropy, we regularize the operator in the following way. First, we diagonalize the time-evolved operator $\mathcal{O}_n$ as
\begin{equation}
	\mathcal{O}_n=V_nD_n{V_n}^{\dagger},
\end{equation}
where $V_n$ is the unitary matrix that diagonalizes $\mathcal{O}_n$. In the second step, we take the absolute value of the eigenvalues of $D_n$ and divide it by its trace to get $\tilde{D}_n=|D_n|/\tr(|D_n|)$. We then construct a positive operator with a unit trace that behaves as a density matrix while keeping the eigenvectors of the observable $\mathcal{O}_n$ intact as
\begin{equation}
	\rho_{\mathcal{O}_n}=V_n\tilde{D}_n{V_n}^{\dagger}.
\end{equation}
Now we can calculate the quantum relative entropy between the two operators $\mathcal{O}_n$ and $\mathcal{O}_n'$ as \begin{equation}
	\mathcal{D}_{KL}(\rho_{\mathcal{O}_n}||\rho_{\mathcal{O}'_n})=\tr({\rho_{\mathcal{O}_n}}(\log\rho_{\mathcal{O}_n}-\log\rho_{\mathcal{O}'_n})).
	\label{relent}
\end{equation}
where $\rho_{\mathcal{O}_n}$ and $\rho_{\mathcal{O}'_n}$ are positive operators of the unit trace obtained from the regularization of operators $\mathcal{O}_n$ and $\mathcal{O}'_n$ respectively. As depicted in Fig. \ref{fig:fid1}c, the distance between the two operators increases rapidly when the level of chaos is lower in the dynamics. This indicates that the operator becomes less prone to error in the Hamiltonian with increasing level of chaos. Ultimately, this makes quantum state tomography more immune to error in the presence of chaos, as already demonstrated in Fig. \ref{fig:fid1}a.

To further elucidate the decline in the rate of reconstruction fidelity, we connect the operator incompatibility to the information gain. We quantify the incompatibility of two operators $\mathcal{O}_n$ and $\mathcal{O}'_n$, as defined above, with time as
$\mathcal{I}_{\mathcal{O}}(t_n)=\frac{1}{2j^4} \tr(|[\mathcal{O}_n,\mathcal{O}'_n]|^2)$. We define this as the error OTOC as   $\mathcal{I}_{\mathcal{O}}$  can be realized as a quantity like the standard OTOC evolved under an effective \emph{error unitary} $\mathcal{U}_{n}=U'^{n}_{\tau}U^{\dagger n}_{\tau}$
\begin{equation}
	\mathcal{I}_{\mathcal{O}}(t_n)=\frac{1}{2j^4} \tr(|[\mathcal{O},\mathcal{U}^{\dagger}_n\mathcal{O}\mathcal{U}_n]|^2). 
	\label{error_otoc}
\end{equation}
The above expressions employs the Hermiticity property of the physical observables and the cyclic property of the trace  to obtain (see supplementary section for details).

This form of error scrambling is very general for dynamics with time-dependent as well as time-independent Hamiltonian.
 If the error enters through the Hamiltonian, then the error unitary for time-independent Hamiltonian becomes $\mathcal{U}(t)=e^{iH't}e^{-iHt}$ and for time-dependent kicked Hamiltonian (e.g. kicked top) $\mathcal{U}_{n}=U'^{n}_{\tau}U^{\dagger n}_{\tau}$, where  $U_{\tau}$ is the Floquet map. 
 The expression, $[\mathcal{O},\mathcal{U}^{\dagger}(t)\mathcal{O}\mathcal{U}(t)]$, in the presence of zero errors gives zero as $\mathcal{U}(t)=e^{iH't}e^{-iHt}$ reduces to the identity. 
Hence, it is indeed ``error scrambling" that gets activated only in the presence of errors. The ``scrambling part" comes from the natural structure of the usual OTOCs. In this supplementary material, we provide a qualitative understanding of these signatures. 

The growth of OTOC has been studied extensively as a quantifier for  information scrambling under chaotic dynamics~\cite{maldacena2016bound,swingle2016measuring,
	hashimoto2017out,kukuljan2017weak,swingle2018unscrambling,wang2021quantum,sreeram2021out,varikuti2022out}. 
Similarly, growth of $\mathcal{I}_{\mathcal{O}}$ implies \emph{scrambling of errors} with time. It is apparent from Fig. \ref{fig:fid1}d that the rate of error scrambling decreases with an increase in the value of the chaoticity parameter $\lambda$. This signifies that the measurement record is less affected by the error in the dynamics when one approaches a greater extent of chaos. {In Eq. (\ref{rml2}), the measurement record $\bf M'$ is obtained from the \emph{true} (perturbed) dynamics, but the covariance matrix $\bf C$, and $\tilde{\mathcal{O}}$ are determined from the experimentalist's version of the dynamics (ideal or unperturbed). Thus, a higher rate of error scrambling for regular dynamics leads to a faster decay of reconstruction fidelity as the measurement record is more vulnerable. How errors scramble across a chaotic system, as given by Eq. (\ref{error_otoc}), is itself an interesting quantifier of quantum chaos.
	Here, we notice the correlation between scrambling of errors as captured by the incompatibility between the operator and its time evolution through the \textit{error unitary}  in  Eq. (\ref{error_otoc}) and OLE, as viewed from the lens of quantum tomography under chaotic dynamics. This links two fundamental quantifiers of quantum chaos, complements findings in \cite{zurek2020information, bhattacharyya2022towards} and provides a different but more intuitive connection. 
	
    {In previous works,   \cite{zurek2020information, bhattacharyya2022towards} , one divides a joint many-body system Hilbert space into two parts and then considers two local operators with support on these parts. The central idea then is to take a Haar average over the set of all unitaries on the two fixed subsystems and compute the OTOC, which is related to a specific kind of {LE} where the Hamiltonian of the larger subsystem serves as the unperturbed Hamiltonian and the perturbation arises out of the interaction between two subsystems. We, however,  do not have to rely on special operators on disjoint subspaces. After all,  OTOCs also capture the incompatibility of the same operator with its Heisenberg evolved avatar at a future time. Secondly, in contrast to these studies, we connect the OLE to how fast the errors scramble due to chaos, as captured by the error OTOCs, and deal with the total system Hamiltonian and perturbations to it.}
	{The previous results involve averaging, these are  (or equivalently will involve the application of typical Haar random unitary on each subsystem) statistical in nature. This is a very elegant result, however, in our  opinion the concepts of chaos and LE are valid for individual trajectories and quantify how much a single trajectory departs from its course  over time in single shot experiments. The LE and chaos form the cornerstone of classical statistical mechanics \cite{pathria2017statistical} to address questions like the basis of ergodic hypothesis and irreversibility of the second law, and therefore to make connections independent of statistical arguments is conceptually satisfying.}
	

    It is surprising that the observables are less sensitive to errors when the
dynamics are more chaotic. This runs counter to the general expectation that
chaos leads to greater scrambling. In previous works the LE illustrated the stability of quantum evolutions in systems like the kicked top and the coupled rotator\cite{peres1984stability, peres1997quantum}. In these works by Asher Peres, the time evolution of the same initial quantum state for two slightly different values of the kicking strength in the kicked top Hamiltonian is considered \cite{peres1997quantum}. After certain number of time steps, the overlap between the evolved states drops to a lower value if the initial spin coherent state is located in a chaotic region of the classical phase space. However, the two evolved states remain close to each other if the initial spin coherent state is centered near a stable fixed point. Later, LE has been studied rigorously, theoretically, and experimentally for many complex quantum systems \cite{gorin2006dynamics, Goussev2012loschmidt}. It is shown that for typical/random quantum states, the fidelity decay is much faster for regular dynamics than chaotic dynamics, which is entirely opposite to Peres's conclusion as he had considered a very special initial coherent state. Thus, for random quantum states, the quantum chaotic dynamics is less prone to error than for regular dynamics. {Similarly, our work connects the theory for OLE by connecting it to continuous weak measurement tomography and the fidelities obtained, and we show that the random observables are less prone to error when the dynamics is more chaotic.}
	
\section{Discussion}
Classically, if we know the dynamics exactly, then as we maintain a constant coarse-grained tracking of the trajectory, we gain exponentially fine grained information about the initial condition. In the quantum setting, this means as we keep track of the measurement record with fixed signal-to-noise, we gain increasing information about the initial condition. Now, what happens to this information gain in the presence of {systematic} errors? This has not been answered either in classical or quantum chaos literature. In this work, we address this very question.

	We find dynamical signatures of chaos that quantify the scrambling of {systematic} errors across a many-body quantum system that has consequences on the performance of quantum information and simulation protocols. We also give an operational interpretation of the OLE by connecting it to the growth of distance between operators evolved in continuous measurement quantum tomography. 
    More importantly, we identify a signature of chaos that is intimately connected with the very notion of chaos -- randomness, errors, predictability, and information. {Using the  kicked top, we have studied how for a fixed family of maps the information gain in tomography in the presence of errors is related to the degree of chaos. It will certainly be interesting to see how far these ideas go towards true universality.}

{While our work deals with systematic errors}, it   paves the way for further studies in the performance of quantum simulations under inadvertent noise.
	In the era of noisy, intermediate-scale quantum (NISQ) devices~\cite{preskill2018quantum}, the accuracy of an analog quantum simulator will decay after just a few time steps. The reliability of such analog quantum simulators is highly questionable even for state-of-the-art architecture when it is likely to exhibit quantum chaos~\cite{hauke2012can,lysne2020small}.
	On the contrary, the digital quantum simulation is often associated with the inherent Trotter errors~\cite{heyl2019quantum} because of the discretization of the time evolution of a quantum many-body system as a sequence of quantum gates. Thus, a better understanding of errors in simulating many-body quantum systems and information processing protocols that exploit such rich dynamics is paramount.
	These signatures of chaos can be further explored using state-of-the-art experimental techniques involving cold atoms interacting with lasers and magnetic fields \cite{chaudhury2009quantum}.

    We would like to re-emphasize that we have a new way of studying quantum signatures of chaos under errors. Continuous weak measurement tomography and information gain provides us a window  to study both quantum and classical chaos. The characterization of chaos, quantum and classical, on information gain, sensitivity to perturbations  is conceptually satisfying.
    The KS entropy formulation is a fundamental way of studying chaos that has not been done in the presence of errors and has some advantages over the studies involving LE. Any experiments involving the LE involves a backward evolution which is a huge challenge in some experimental setups. Moreover, as discussed earlier, projective measurements are expensive and cumbersome. What we propose is a single shot and, as discussed in the references, an experimentally feasible way of probing quantum chaos that needs no backward evolution of projective measurements.

    The archive of the measurement record, as a continuous stream of information, has the signatures of the underlying dynamics as well as scrambling of errors in that dynamics. This is much akin to the volume of work on time series analysis done in the field of classical chaos. These and similar concepts are making their presence felt in quantum information science with the concept of classical shadows in quantum mechanics being actively pursued. 
    In future work, we hope to further build upon our results to develop quantum analogs of the ``classical shadowing lemma" that guarantee a \emph{true} classical trajectory in the neighbourhood of any arbitrary simulated trajectory of a chaotic system in the presence of truncation errors due to finite precision~\cite{anosov1967geodesic,grebogi1990shadowing,sauer1991rigorous,sauer1997long,vanivcek2004dephasing}.
    Our results linking LE, error scrambling, and OTOCs will be helpful to the condensed matter community as well and in addressing broader issues involving non-integrable quantum systems {as well their applications in quantum limited metrology}~\cite{pandey2020adiabatic, fiderer2018quantum}.

	\section{Supplementary Material}
	
	Supplementary Material includes derivation of Eq. \ref{error_otoc} and a detailed Bayesian view of continuous measurement quantum tomography.  

\section{Acknowledgements}
	
	We are grateful to Arul Lakshminarayan for useful discussions. We thank Sreeram PG for the helpful discussions. We acknowledge BS Datta Vikas for his inputs during the initial stage of this work. The authors would like to thank HPCE, IIT Madras, for providing the computational facility for numerical simulations. This work was supported in part by grant  DST/ICPS/QusT/Theme-3/2019/Q69 and New faculty Seed Grant from IIT Madras. The authors were supported, in part, by a grant from Mphasis to the Centre for Quantum Information, Communication, and Computing (CQuICC) at IIT Madras. The authors acknowledge ANRF MATRICS grant and ANRF NQM (National Quantum Mission). N.D.V. acknowledges funding from the Italian Ministry of University and Research (MUR) through project DYNAMITE QUANTERA2\_00056, in the frame of ERANET COFUND QuantERA II—2021 call co-funded by the European Union (H2020, GA No. 101017733). This work was supported by the Provincia Autonoma di Trento, and Q@TN, the joint lab between University of Trento, FBK—Fondazione Bruno Kessler, INFN—National Institute for Nuclear Physics, and CNR—National Research Council. Views and opinions expressed are however those of the author(s) only and do not necessarily reflect those of the European Union or of the Italian Ministry of University and Research. Neither the European Union nor the granting authority can be held responsible for them.

	\bibliographystyle{unsrt}
	\bibliography{main}    

\appendix
		\section{ Derivation of Error scrambling expression}
			Let the operators at time $t$ be $\mathcal{O}(t)=U^{\dagger}(t)\mathcal{O}U(t)$ and $\mathcal{O'}(t)=U'^{\dagger}(t)\mathcal{O}U'(t)$, where $U(t)$ is the unitary for the unperturbed dynamics and $U'(t)$ is the unitary for perturbed dynamics. Thus, the operator incompatibility is 
		\begin{equation}
			\mathcal{I}_{\mathcal{O}}(t)=\frac{1}{2j^4} \tr(|[\mathcal{O}(t),\mathcal{O}'(t)]|^2).
			\label{opinc}
		\end{equation}
		Now one can simplify this expression using the Hermiticity property of the physical observables $\mathcal{O}(t)=\mathcal{O}^{\dagger}(t)$ and the cyclic property of trace to obtain
		
		\begin{equation}  
			\mathcal{I}_{\mathcal{O}}(t)=\frac{1}{2j^4} \tr(|[\mathcal{O},\mathcal{U}^{\dagger}(t)\mathcal{O}\mathcal{U}(t)]|^2). 
		\end{equation}
		This is the expression for error scrambling, which helps us to connect the operator incompatibility (out-of-time-order correlator) and Loschmidt echo. If the error enters through the Hamiltonian, then the error unitary for time-independent Hamiltonian becomes $\mathcal{U}(t)=e^{iH't}e^{-iHt}$ and for time-dependent kicked Hamiltonian (e.g. kicked top) $\mathcal{U}_{n}=U'^{n}_{\tau}U^{\dagger n}_{\tau}$, where  $U_{\tau}$ is the Floquet map. 
 The expression, $[\mathcal{O},\mathcal{U}^{\dagger}(t)\mathcal{O}\mathcal{U}(t)]$, in the presence of zero errors gives zero as $\mathcal{U}(t)=e^{iH't}e^{-iHt}$ reduces to the identity. 
Hence, it is indeed ``error scrambling" that gets activated only in the presence of errors. The ``scrambling part" comes from the natural structure of the usual OTOCs.
		
\section{Understanding  information gain in  tomography with errors}
		Here, we describe information gain and the general behavior of reconstruction fidelity as shown in the main text.
		Later in this section, we will demonstrate the effect of the magnitude of perturbation on the fidelity obtained in tomography. 
		We observe that independent of the degree of chaos in the dynamics, the fidelity initially rises despite errors and then starts to decline after attaining a peak. As described in the main text, we consider the density matrix of Hilbert-space dimension $d$ that can be realized as a generalized Bloch vector $\bf r$ by expanding $\rho_0=I/d+\Sigma^{d^2-1}_{\alpha=1}\ r_\alpha E_\alpha$ in an orthonormal basis of traceless Hermitian operators $\{E_\alpha\}$.

		The probability of reconstructing a state $\rho_0$ is \cite{sahu2022effect} 
		\begin{equation}
			p({\bf \rho_0|M, \mathcal{L}, \mathcal{M}}) = A \: p({\bf M|\rho_0, \mathcal{L}, \mathcal{M}})\: p(\rho_0|\mathcal{L}, \mathcal{M})\: p(\mathcal{L}, \mathcal{M}),
			\label{tom_eq}
		\end{equation}
		where $A$ is a normalization constant. The first term $p({\bf M|\rho_0, \mathcal{L}, \mathcal{M}})$, is the probability of acquiring a measurement record $\bf M$, given an initial state $\rho_0$, the dynamics $\mathcal{L}$ (choice of unitaries), and the measurement process $\mathcal{M}$ (choice of operators $\mathcal{O}$ for generating measurement record). This term contains the errors due to shot noise and helps one to quantify the signal-to-noise ratio in various directions in the operator space independent of the state to be estimated. Thus, in the limit of negligible backaction $p({\bf M|\rho_0, \mathcal{L}, \mathcal{M}})$ is identical to the probability distribution corresponding to the measurement history $\bf M$ for a given Bloch vector $\bf r$ \cite{silberfarb2005quantum, smith2006efficient, merkel2010random, madhok2014information, sreeram2021quantum},
		 
		\begin{figure}
			\centering
			\includegraphics[width=6.35cm, height=6.3cm]{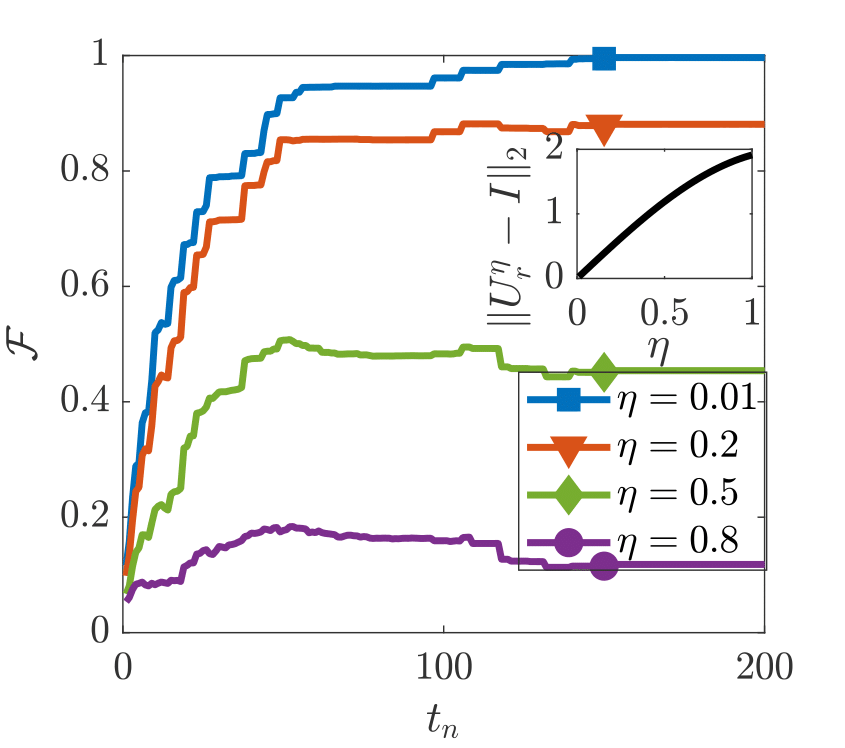}  	
			\caption {{Reconstruction fidelity as a function of time in the limit of vanishing shot noise for an increase in the perturbation of $\{E_\alpha\}$. The measurement operators are the perturbed ordered set \{$E'_1$, $E'_2$, ..., $E'_k$\}  with ordered Bloch vector components of the initial state $\rho_0$ (i.e. that corresponds to the Bloch vector components, $r_{\alpha} = \tr[\rho_0 E_{\alpha}]$  in a particular order of their magnitudes). The perturbed operators $\{E'_\alpha\}$ are generated by applying a fractional power $\eta$ of a random unitary $U_r$. The inset figure shows the Euclidean norm of the difference between $U^{\eta}_r$ and Identity $I$, which increases with an increase in the value $\eta$.}}
			\label{fig:Blochp}
		\end{figure}
		
		\begin{equation} 
			\begin{split}
				p({\bf M|r}) & \varpropto \mathrm{exp}\ \Big\{-\frac {N^2}{2\sigma^2}\sum_{i}[M_i-\sum_{\alpha}\tilde{\mathcal{O}}_{i\alpha}r_\alpha]^2\Big\}
				\\
				& \varpropto \mathrm{exp}\ \Big\{-{\frac {N^2}{2\sigma^2}\sum_{\alpha,\beta}({\bf r-r_{ML}})_\alpha\ C^{-1}_{\alpha\beta}\ ({\bf r-r_{ML}})_\beta\Big\}}.
				\label{pdf_uniform_prior_supp}
			\end{split}
		\end{equation}
		Therefore, this term estimates the information gained, given a density matrix, in different directions in the operator space. Now we have the second term $p(\rho_0|\mathcal{L}, \mathcal{M})$, which is the posterior probability distribution relating the knowledge of the dynamics and the measurement operators. On that account, in the limit of vanishing shot noise and with complete knowledge of the system dynamics for given measurement observables $\{E_{\alpha}\}$, this conditional probability is continuously updated and ultimately becomes a product of Dirac-delta functions. Once we obtain an informationally complete measurement record, each Dirac-delta function identifies a particular Bloch vector component. The term $p(\mathcal{L}, \mathcal{M})$ in Eq. (\ref{tom_eq}) can be absorbed in the constant as it gives the prior information about the choice of dynamics and measurement operators. Thus, Eq. (\ref{tom_eq}) separates the probability of quantum state estimation into a product of two terms (up to a constant) \cite{sahu2022effect}.
		\begin{widetext}
			\begin{equation} 
				\begin{split}
					p({\bf \rho_0|M, \mathcal{L}, \mathcal{M}}) &\varpropto \mathrm{exp}\ \Big\{-\frac {N^2}{2\sigma^2}\sum_{i}[M_i-\sum_{\alpha}\mathcal{O}_{i\alpha}r_\alpha]^2\Big\}\      p(\rho_0|\mathcal{L}, \mathcal{M})  \\
					&\varpropto  \mathrm{exp}\ \Big\{-{\frac {N^2}{2\sigma^2}\sum_{\alpha,\beta}({\bf r-r_{ML}})_\alpha\ C^{-1}_{\alpha\beta}\ ({\bf r-r_{ML}})_\beta\Big\}}\ p(\rho_0|\mathcal{L}, \mathcal{M}) 
				\end{split}
			\end{equation}
		\end{widetext}

		In the limit of zero shot-noise, the errors due to the first term are zero, and we may purely focus on the conditional probability distribution, $p(\rho_0|\mathcal{L}, \mathcal{M})$. In terms of the observables in continuous measurement tomography,  one can express $p(\rho_0|\mathcal{L}, \mathcal{M}) = p( \bf r| \mathcal{O}_1, \mathcal{O}_2, ..., \mathcal{O}_n )$,  giving the conditional probability of the density matrix parameters $\bf r$ till the time step $n$. For example, consider  the measurement operator at the first $k$ time steps are the ordered set \{$E_1$, $E_2$, ..., $E_k$\}, giving precise information about Bloch vector components \{$r_1$, $r_2$, ..., $r_k$\}. The conditional probability distribution at time $k$ is,

		\begin{widetext}
			\begin{equation}
				\begin{split}
					p(  \textbf{r}| E_1,E_2, ..., E_k ) = \delta(r_1-\tr[E_1 \rho_0])\ \delta(r_2-\tr[E_2 \rho_0])\ ...\ \delta(r_k - \tr[E_k \rho_0])\ \delta \big(\sum^{d^2-1}_{\alpha \neq 1, 2,...k}r_{\alpha}^2 = 1 - 1/d - r_{1}^2 -  r_{2}^2 ...- r_{k}^2 \big).
				\end{split}
			\end{equation}
		\end{widetext}

		Each noiseless measurement above gives us complete information in one of the orthogonal directions. For example, after the first measurement,
		\begin{equation}
			p(\textbf{r}| E_1) =  \delta(r_1-\tr[E_1 \rho_0])\ \delta \big(\sum^{d^2-1}_{\alpha \neq 1}r_{\alpha}^2 = 1 - 1/d - r_{1}^2 \big).
		\end{equation}
		Therefore, once $r_1$ is determined, the rest of the $d^2 - 2 $ Bloch vector components are constrained to reside on a surface given by the equation $\sum^{d^2-1}_{\alpha \neq 1}r_{\alpha}^2 = 1 - 1/d - r_{1}^2 .$  
		The state estimation procedure shall select a state based on incomplete information
		consistent with $r_1$ as determined precisely by the first measurement and the remaining Bloch vector components from a point on this surface. Therefore, qualitatively, the average fidelity of the estimated state is proportional to the area of this surface. After $k$ time steps, the error is proportional to the area of the surface consistent with the equation  $\sum^{d^2-1}_{\alpha \neq 1}r_{\alpha}^2 = 1 - 1/d - r_{i}^2 -  r_{j}^2- ... - r_{k}^2$. This area, quantifying the average error, decreases with each subsequent measurement. 
		\begin{figure}
			\centering
			\includegraphics[width=6.37cm, height=6.3cm]{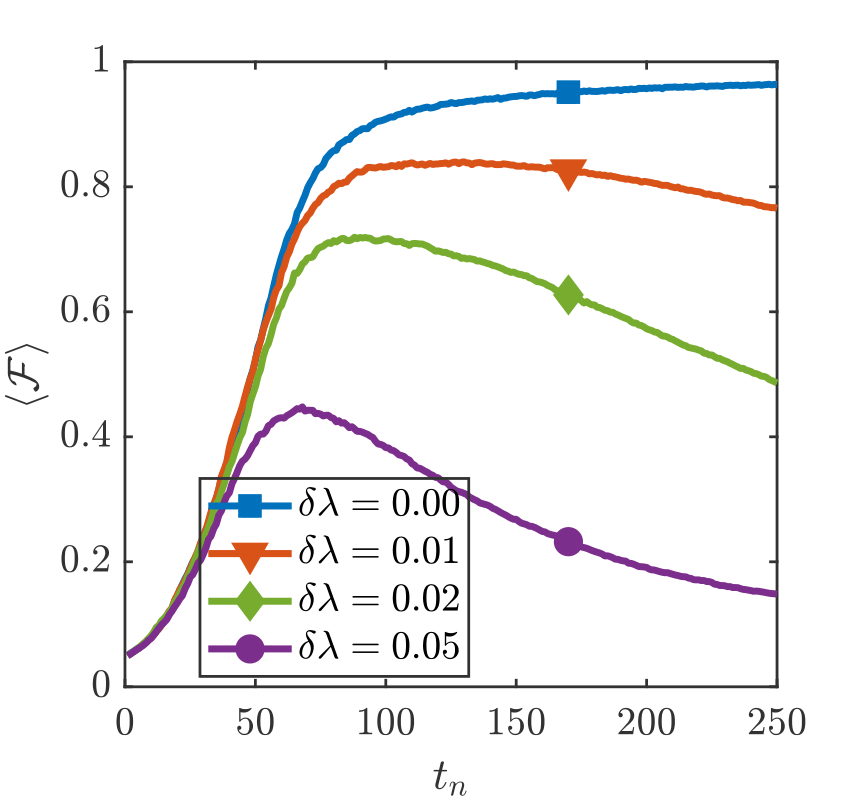}  	
			\caption {Reconstruction fidelity as a function of time for an increase in perturbation strength. The measurement record is generated for spin $j=10$. Here we consider rotation angle $\alpha=1.4$ and kicking strength $\lambda=7.0$ for the quantum kicked top.}
			\label{fig:fidp}
		\end{figure}
		
		{To see it in another way, consider the fidelity between the actual and reconstructed state.} The fidelity $\mathcal{F}=\bra{\psi_0}\bar{\rho}\ket{\psi_0}$,
		\begin{equation}
			\mathcal{F} = 1/d+\Sigma^{d^2-1}_{\alpha=1}\ \bar{r}_\alpha {r}_\alpha
		\end{equation}
		Here ${r}_\alpha$ and $\bar{r}_\alpha$ are the  Bloch vectors for ${\rho_0}$ and $\bar{\rho}$ respectively. 
		As one makes measurements, ${E_{1}}$, ${E_{2}}$, ..., ${E_{k}}$ and gets information about the corresponding Bloch vector components (with absolute certainty in the case of zero noise for example), one can express the fidelity as 
		
		\begin{equation}
			\mathcal{F} = 1/d+\Sigma^{k}_{i=1}\  {r}_i^2 + \Sigma^{d^2-1}_{\alpha \neq 1, 2,...k}\ \bar{r}_\alpha {r}_\alpha 
			\label{unpert_fid}
		\end{equation}
		The term $ \Sigma^{k}_{i=1}\  {r}_i^2$ puts a lower bound on the fidelity obtained after $k$
		measurements. Here $\bar{r}_{\alpha}$ for ${\alpha \neq 1, 2, ..., k}$ represent the state estimator's 
		guess for the unmeasured Bloch vector components consistent with the constraint $\big(\sum^{d^2-1}_{\alpha \neq 1, 2,..., k}{r}_{\alpha}^2 = 1 - 1/d - {r}_{1}^2 -  {r}_{2}^2 ...- {r}_{k}^2 \big)$.
		It is this guess that picks a point from the surface with an area consistent with the above constraint.
		
		Consider the same scenario but now with perturbations to the system dynamics. Once we add errors to the dynamics, this problem becomes richer. In the presence of errors, not only does the dynamics generate a rate of information gain but the same dynamics can be potentially sensitive to errors depending upon the degree of chaos in the system. How does the tension between information gain as well as the scrambling of errors, both due to chaos, influence the fidelity of reconstruction is one key contribution of our work. In order to make these statements about the system dynamics, spread of errors and reconstruction that are independent {of specific properties of the initial state to be reconstructed, and the choice of operators we have considered random states and random initial observables.}

  The estimate of the density matrix gets modified as
		
		\begin{widetext}
			\begin{equation}
				\begin{split}
					p(  \textbf{r}| E'_1, E'_2, ..., E'_k ) = \delta(r'_1-\tr[E'_1 \rho_0])\ \delta(r'_2-\tr[E'_2 \rho_0])\ ...\ \delta(r'_k - \tr[E'_k \rho_0])\ \delta \big(\sum^{d^2-1}_{\alpha \neq 1, 2,...k}{r'}_{\alpha}^2 = 1 - 1/d - {r'}_{1}^2 -  {r'}_{2}^2 ...- {r'}_{k}^2 \big).
				\end{split}
			\end{equation}
		\end{widetext}
		{Here, $E'_1, E'_2, ..., E'_k$ are the perturbed operators leading to a slightly inaccurate estimate of the Bloch vector components $r'_1, r'_2, ..., r'_k$ respectively. The operators $\{E'_\alpha\}$ are obtained by rotating $\{E_{\alpha}\}$ by a unitary $U^{\eta}_r$, where $U_r$ is a random unitary and $\eta$ is a fractional power which makes $U^{\eta}_r$ close to identity. The Euclidean norm of the operator $U^{\eta}_r-I$ is less when $\eta$ is small. Thus, $\eta$ serves as the strength of perturbation in this analysis of Bloch vector components.}
		
		Despite the perturbation, the uncertainty of the Bloch vector components $r_{\alpha}$ for $\alpha\neq 1,2,...,k$ reduces to the area of the surface consistent with the equation $\sum^{d^2-1}_{\alpha \neq 1, 2,...k}{r'}_{\alpha}^2 = 1 - 1/d - {r'}_{1}^2 -  {r'}_{2}^2 ...- {r'}_{k}^2$. The fidelity between the original and the estimated state now reads as 
		\begin{equation}
			\mathcal{F} = 1/d+\Sigma^{k}_{i=1}\  {r}'_i{r}_i + \Sigma^{d^2-1}_{\alpha \neq 1, 2,...k}\ \bar{r}_\alpha {r}_\alpha, 
			\label{pert_fid}
		\end{equation}
		that we can see in Fig. \ref{fig:Blochp}.
		We know that the overlap between two Bloch vectors is maximum only when they are exactly aligned in the same direction, and the overlap decreases when they move far from each other. Comparing the second terms of Eq. (\ref{unpert_fid}) and Eq. (\ref{pert_fid}) it is now clear why with an increase in perturbation, the initial rise in fidelity is less. 
		Therefore, the drop in fidelity is more, and the fidelity saturates at a lower value if the perturbation is more, as illustrated in Fig. \ref{fig:fidp}.
		For relatively weaker perturbations, the fidelity will continue to increase when there is an information gain despite such errors to the measurement operators. The partially inaccurate information about the $j$th Bloch vector owing to perturbations to the dynamics still offsets the estimator's guess 
		of the Bloch components of the unmeasured $j$th direction in the operator space determined by $E_j$.


\end{document}